\documentclass[pre,twocolumn,aps,amssymb,footinbib,showpacs]{revtex4}
\usepackage{bm}
\usepackage{amssymb}
\usepackage{times}
\usepackage{latexsym}
\usepackage{graphicx}
\usepackage{color}
\usepackage{subfigure}
\newcommand {\beq}{\begin{equation}}
\newcommand{\eeq}{\end{equation}}
\newcommand {\bea}{\begin{eqnarray}}
\newcommand{\eea}{\end{eqnarray}}
\begin{document}
\title{Solitons in a hard-core bosonic system: Gross-Pitaevskii 
type and beyond}
\author{ Radha Balakrishnan$^{1}$ and  Indubala I Satija$^{2}$}
\affiliation{$^{1}$ The Institute of Mathematical Sciences, 
Chennai 600113, India}
\affiliation{$^{2}$ School of Physics, Astronomy and 
Computational Sciences, George Mason University, Fairfax, 
VA 22030, USA}
\begin{abstract}
A unified formulation that obtains  
solitary waves  for various background densities
in the Bose-Einstein condensate  
of a system of hard-core bosons with nearest neighbor attractive
interactions is presented.   
 In general,
two species of solitons appear:
A nonpersistent (NP) type  that fully
delocalizes at its maximum speed, and a persistent (P) type that
survives  even at its  maximum speed, and transforms into a 
periodic train of solitons
above this speed. When the  background condensate density 
is nonzero, both species coexist, the soliton is associated with a 
constant intrinsic
frequency, and its maximum speed is the speed of sound. 
In contrast, when the background condensate density is zero,
the system has neither a fixed frequency, nor a speed
of sound. Here, the   maximum soliton speed
depends on the frequency, which can be tuned to lead to 
a cross-over between the NP-type and the  P-type
at a certain critical frequency, determined by the energy parameters
of the system.  
We provide a single functional form for the soliton profile,
from which diverse characteristics for 
various background densities  can be obtained.  Using the
mapping  to spin systems enables us 
to characterize  the corresponding class of magnetic solitons in 
 Heisenberg spin chains with different types of anisotropy,
in a unified fashion.

\end{abstract}
\maketitle
\section{Introduction}

Experimental demonstration \cite{expt,Bexpt, becker,dutton, burger,frantz} 
of solitary waves/solitons \cite {def} 
in Bose-Einstein condensates (BEC)\cite{pethick, gpe}  is one of the 
hallmarks of quantum coherence
inherent in ultracold atomic systems
As predicted theoretically in the Gross-Pitaevskii 
equation (GPE) \cite {gpe},
which describes  {\em weakly interacting} bosons in the mean 
field approximation, a condensate of Rb atoms with repulsive
interactions  was found to support dark solitary 
waves (density depressions) \cite{expt}, 
while  a Li condensate \cite{Bexpt}
with attractive interactions supported bright 
solitary waves (density elevations)\cite{daux}.
Various recent theoretical studies \cite{mish} have investigated
 soliton evolution in quantum many body systems
 to understand   
the role of quantum fluctuations on mean field solutions.
Intrinsically nonlinear, the BEC systems continue to remain
 an active area to explore the presence of  
 nonlinear localized modes.
In view of the fact that GPE also describes nonlinear optical 
systems, these studies are relevant beyond the BEC literature.
 
In  our previous work \cite {ourprl} , we investigated 
the propagation of solitonic excitations in a  
system of  hard-core bosons (HCB),
which describes  {\it strongly repulsive} 
bosons. By mapping an extended Bose-Hubbard model \cite{sach}
for hard-core bosons on a lattice, with  nearest 
neighbor (nn) hopping energy $t$  and  (nn) 
interactions $V$ 
 to a spin model, we used
spin-coherent states \cite{radcliffe} to  obtain the condensate 
density for HCB 
as $\rho^s = \rho(1 - \rho)$, where $\rho$ is the bosonic (particle) 
density for the HCB system.  We  derived the continuum evolution equation
for the condensate wave function, 
which we called HGPE, `H' standing for HCB. 
 The only model-dependent  effective energy  parameter 
that appears in HGPE is $E_{e} = (t - V)/t$.
 
For the case $E_e > 0$, we analyzed unidirectional solitary wave 
excitations in HGPE  when the 
background density $\rho_0$
contains  {\it both particles and holes},  For a hard-core
system, this implies $0 < \rho_0 < 1$. We refer to this as
the  `fractional filling' case.  This corresponds to 
a non-zero condensate density  $\rho^{s}_0$ in the 
background. 
 Under these conditions, the  system was shown to possess
 an intrinsic  speed of sound and an intrinsic 
 frequency parameter.  These are
{\it fixed} in the sense that they depend on the 
given background density and the system parameters.
This frequency can be shown to
be  just the frequency associated with the 
phase of the homogeneous condensate in the background.

At half-filling,  both bright and dark 
solitons which are mirror images of each other are
 supported for the density $\rho$. Further, both are 
{\em nonpersistent} (NP)
  type solitons that 
  flatten out and delocalize at their maximum speed, 
which is the speed of sound in the system.
 Intriguingly, for half-filling, the behavior of  solitons for 
the condensate density $\rho^s$  in
 this strongly repulsive system can be shown to be \cite{pla}
very similar  to that of the  GP  soliton 
in  a weakly repulsive system, since 
 it is dark,  and  delocalizes at  sonic  speed.

Away from half-filling, we found {\em two  distinct species}  
of  solitons  that  coexist  in the HCB system. 
For $0 < \rho_0 <  1/2$~~($1 > \rho_0 > 1/2$),
one is a  NP-type   dark (bright) soliton for density, that
delocalizes completely at the
  speed of sound,  while the other is a novel  {\em persistent}
 P-type  bright (dark) soliton that  survives  as a localized entity, 
 even at its maximum speed, 
 In addition, 
the  P-soliton  transforms 
into a train of solitons at supersonic speeds, quite unlike a GP soliton. 
The corresponding condensate density soliton  for the NP-type is  always
 dark, while that for the P-type not only survives at the speed of sound,  
but also becomes completely bright at this speed \cite{ourprl}. 
This brightness is quite unexpected
in a very strongly repulsive system like the HCB. Preliminary results on 
collision of these solitary waves \cite{arxiv} show that 
they emerge unscathed, suggesting that they
could be strict solitons.
Additionally,  
in a recent study \cite{DMRG}, we have  also shown  that
 both the above species of solitary 
waves remain stable on the lattice under time evolution, 
and also survive quantum fluctuations, for experimentally 
accessible time scales.

\begin{figure}[htbp]
\includegraphics[width = 1.1\linewidth,height=1.1
\linewidth]{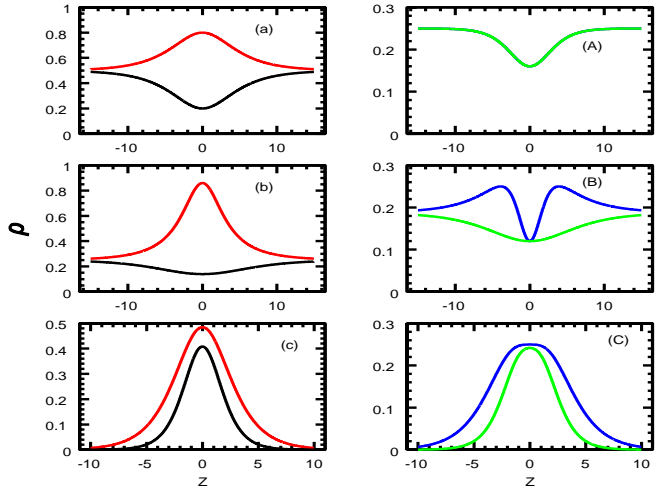}
\leavevmode \caption{(color on line) Solitary waves with 
background density
 $\rho_0= 0.5$ (a), 
$\rho_0= 0.25$ (b), and $\rho_0=0$ (c) for fixed 
$E_e = 0.1$ and $v/c = 0.8$.
Here red and black show the persistent (P-type) and 
non-persistent (NP-type) 
solitons. (A), (B) and (C) show the corresponding condensate 
density $\rho^{s}$, where the blue and the green respectively 
correspond to P and NP-type, respectively.
For $\rho_0= 0.5$, the plots for $\rho^{s}$  for the two types 
overlay each other. For $\rho_0=0$ which exhibits only bright solitons, 
the P and the NP
types plotted correspond to $E_{\omega}= 0.15$ and $0.3$ respectively. }
\label{Sall}
\end{figure}

  A natural question that arises is whether the HCB system 
  can  support solitary wave excitations  for $E_e > 0$, 
when the background density has  {\it only}  particles  
or {\it only} holes,   
 i.e., for $\rho_0 =1$ or $\rho_0 = 0$. We  shall refer 
to this as the  ``integer filling " case. This corresponds to 
a vanishing condensate density in the background. This
represents a distinct class 
 in contrast to the   fractional filling case, because this system has 
neither  an  intrinsic speed of sound to limit
the soliton speed,   nor  a 
fixed intrinsic frequency. In addition, it is also worth exploring the 
existence of solitons for
 $E_e \le 0$.

In this paper, we address above questions within a
{\em unified formulation}  that obtains a single functional 
form for the density soliton which  is valid 
 for  {\em both}  fractional filling and integer filling 
background densities,
as well as for $E_e >0$ and  $E_e \le 0$. 
Using this, we can obtain the 
diverse  characteristics of solitons for the various possible cases. 
We find that for integer filling,  the  soliton  is 
characterized  both by its speed and 
 by  its  independently  tunable frequency parameter.
 This is unlike the fractional filling case \cite{ourprl}, when 
it is characterized only by its speed. 
 Interestingly,  maximum soliton speed  is seen to depend on this 
frequency $\omega$. leading to two competing energy 
scales, $ E_{\omega} =\hbar \omega/t$ and the effective energy 
$E_{e}$.  

For $E_e >0$, integer fillings, while both NP and P-type  solitons appear,
 they do not  coexist.  Depending upon the relative 
strength of $E_{\omega}$ and $E_e $,
the system supports either  
 a P -soliton that persists at its 
(frequency-dependent) maximum speed  or a NP- soliton 
which delocalizes at this speed. 
Thus the frequency  can be {\em tuned} to lead to 
a cross-over between the NP-type and the  P-type
soliton at a certain {\em critical frequency}, determined by $E_e$.

 For $E_e \le 0$, we show that while  only NP-solitons are supported
 for integer fillings, {\em no} soliton 
solutions exist for fractional filling backgrounds.

One general interesting  feature of  HCB  solitons
for  the integer-filling background  (for all $E_e$)  is   that 
unlike the fractional-filling case where the maximum soliton speed 
is limited by the speed of sound,  
high-speed solitons are possible here, 
since the maximum soliton speed is 
 controlled by the tunable frequency.

Finally, using the mapping of HCB to spins, the single functional  form
we  obtain for the  density soliton also enables us 
to classify the  characteristics of magnetic solitons 
in the isotropic Heisenberg
spin chain as well as  the easy-plane and easy-axis anisotropic
 spin chains,  in a unified fashion.

\section{The extended Bose-Hubbard model}
\subsection {The Model}

As is well known, by loading ultracold bose atoms on
to an  optical lattice \cite{optical}  which is created
using standing waves of laser light, it has become possible
to realize  various  models of condensed matter systems, in the cold atoms lab.
More important, it  is also possible to create lattices of different
dimensions, as well as  {\em tune} the value of the 
parameters in the model, experimentally. This motivates us to
consider  the following Hamiltonian for the  extended  
Bose-Hubbard (BH) model \cite{sach} in $d$ dimensions:
\begin{widetext}
\beq
H=-\sum_{j,a}[t \,(b_j^{\dagger} b_{j+a}+h.c.)+ V n_j n_{j+a}]+\sum_j \big [U n_{j}  
(n_{j}-1) + 2tn_{j} \big ],
\label{BH}
\eeq
\end{widetext}
where $b_j^{\dagger}$ ($b_j$) are  the  normal boson creation   
(annihilation) operators, 
which satisfy  the usual commutation relations  $[b_{j}, b_{l}^{\dagger}] = \delta_{jl}$ and  
$[b_j, n_l] = b_{j} \delta_{jl}$, where
$n_{j}$  is the  number operator at site $j$.  $a$ 
 labels nearest-neighbor (nn) sites,
$t$ is the  nn hopping parameter,  $U$  denotes the on-site repulsive energy and 
$V$ is the  nn interaction.
The on-site term  $2tn_j$  has been added so as to 
obtain the correct kinetic energy term 
in the continuum version of the many-body bosonic Hamiltonian,
which will also enable direct comparison with the usual form 
of the GPE.

While several aspects of  model (\ref{BH}), such as quantum 
phase transitions, phase diagrams, etc.
have been studied \cite{sach},  our interest here is to investigate
 whether this model can  support nonlinear dynamical 
excitations like solitons.

\subsection{ BEC evolution for weakly repulsive normal bosons and GPE:  
Order parameter evolution using  bosonic coherent states}

Before proceeding to the BEC of strongly repulsive 
bosons, it is instructive to
study the BH model for normal bosons  and its connection to GPE.
The  usual BH  model contains only the on-site  {\em finite} repulsion 
term $U$ and no nn interaction term $V$ in Eq. (\ref{BH}). 
The order parameter 
of a BEC  is conventionally defined as the
expectation value of the boson annihilation operator. It has been argued 
\cite{langer} that the Glauber (or bosonic) coherent- state representation
may be appropriate for computing the expectation value, since
 it is well known that coherent states are most useful in the 
context of quantum many-body systems which display quantum 
effects in macroscopic scales as in a BEC.

Writing down operator evolution equation of $b_j$ using 
$i\hbar (db_j/d\tau) = [b_j, H]$ where $H$ is the usual BH model,
and $\tau$ stands for time, the  boson coherent state 
expectation value of the above operator equation yields 
the order parameter evolution. Its  continuum version can be 
shown to be identical in form  to  
the following Gross-Pitaevskii equation (GPE) 
\cite{gpe} for the time evolution of the BEC order parameter
 for weakly interacting bosons: 
\begin{equation}
i \hbar \psi_{\tau}  +\frac{\hbar^2}{2m} \nabla^2 \psi - 
g |\psi|^{2} \psi = 0,
\label{gpe}
\end{equation}
where we have  identified the hopping parameter 
$t a^{2} = \hbar^{2}/m$  and  
$U = g  =4 \pi \hbar^{2} \bar{a} /m$ . ( Here $\bar{a}$ is the s-wave 
scattering length in an ultracold dilute bose gas with 
local (weak)  interaction $g = U$ ).
 Since $U > 0$ in Eq. (\ref{BH}),  
 Eq. (\ref{gpe}) describes the condensate dynamics for  weak repulsive 
interaction between bosons. It is to be noted that 
the use of boson coherent state
expectation values  leads to  a mean field description, so that
for the GPE, the condensate density $\rho^{s}$ is equal to 
 the particle density $\rho$.  

Although soliton solutions of GPE  are well known,
 our analysis  presented below  differs from those discussed
in the BEC soliton literature \cite{gpe, daux}. As we shall see,
our systematic  methodology  for obtaining solitons in this  weakly
repulsive GPE  will also be useful  in arriving at a unified formulation 
for finding  soliton solutions for BEC in the strongly repulsive system
described by hard-core bosons, for various background densities.

It is well known that linear modes of the GPE can be found by analyzing 
small amplitude traveling wave solutions of Eq. (\ref{gpe}) to yield the
Bogoliubov dispersion relation \cite{gpe}, which 
shows that the modes are sound waves
with speed $ c_g = \sqrt \frac {g \rho_0}{m}$.
In order to study solitary waves in the $x$-direction, we first set
$\psi = \sqrt{\rho(x , \tau)} \exp {i\phi (x, \tau)}$ 
 in Eq. (\ref{gpe}),
and separate its real and imaginary parts to obtain coupled 
equations for $\rho$ and $\phi$.\\
 To study soliton propagation, typically one  looks for 
unidirectional traveling wave solutions of the form 
\beq
\rho(x,\tau)=\rho_0+f(z) ;\,\,\,
\phi(x,\tau)=\omega \tau + \phi(z),
\label{ansatz}
\eeq
where $z=(x-v\tau)$.
Here, $v$ is the speed of the traveling wave 
and $\omega$ is a 
frequency  parameter.

Using Eqs. (\ref{ansatz})  in the coupled equations for 
$\rho$ and $\phi$ obtained
from Eq. (\ref{gpe}), a lengthy but straightforward  
calculation yields
\begin{widetext}
\beq
-v \rho_{z}+\frac{\hbar}{m} \frac{d}{dz} (\rho \phi_z)  =  0\\
\label{rhoe}
\eeq
\beq
-4 \hbar \rho^{2} \omega + 4 \hbar \rho^{2} v \phi_{z} + 
(\hbar^{2}/m) \rho \rho_{zz} -(\hbar^{2}/2m) 
\rho_{z}^{2} - (2\hbar^{2}/m) \rho^{2}\phi_{z}^2
- 4 g \rho^{3} = 0
\label{phie}
\eeq
\end{widetext}
where the subscript $z$ stands for derivative with respect to $z$.

We are interested in soliton solutions with boundary
conditions  $\rho(z) \rightarrow \rho_0$
and  the derivative $\phi_{z}
\rightarrow 0 $, as $|z| \rightarrow \infty$.
With this,  Eq. (\ref{rhoe}) can be easily integrated  
 to yield
\begin{equation}
\phi_{z}=\frac{mv}{\hbar} \frac{(\rho-\rho_0)}{\rho}.
\label{gradphi}
\end{equation}
Substituting $\phi_z$ from (\ref{gradphi}) into 
Eq. (\ref{phie}),  multiplying by $ \rho_{z} ~\rho^-{2}$ 
and collecting terms appropriately, it becomes possible to integrate
the resulting equation to yield

\begin{equation}
\frac {\hbar^{2}}{2m} \rho_{z}^2 = 
2g  \rho^3 - (2mv^2 - 4 \hbar \omega) 
\rho^2- \lambda_{g} \rho - 2 m v^2 \rho_{0}^2,
\label{cubicg}
\end{equation}
where $\lambda_{g}$ is a constant of integration
to be determined consistently.
The  subscript $g$ is 
used to indicate that the quantity concerned  
corresponds to the GPE case.
  
 Note that the right hand side of the above equation 
is a cubic polynomial. 
Since we are interested in finding
localized solutions for $\rho (z)$, with the asymptotic
boundary condition $ (d\rho/dz) =0$ as $\rho \rightarrow \rho_0$,
we can write
Eq. (\ref {cubicg}) in the form
 
\begin{equation}
\frac{\hbar^{2}}{2m} \rho_{z}^2 = (\rho-\rho_0)^2 [M_g \rho + N_g].
\label{gperhoeqn}
\end{equation}
The unknown quantities  $M_g$,  $N_g$ and $\lambda_g$ are  found
by equating  the right hand sides of Eqs. (\ref{cubicg}) 
and Eq. (\ref{gperhoeqn}). For $\rho_o \ne 0$, we obtain $ 
M_g  =  2g ;\,\, N_g = -2m v^2 ;\,\, 
\lambda_g = -2 \rho_0 [g \rho_0 + 2 m v^2]$, 
along with
\begin{equation}
\omega = \omega_g = - g \rho_0/\hbar.
\label{gomega}
\end{equation}
Thus the frequency $\omega$ that appears in Eq. (\ref{ansatz})
is a constant for the GP soliton.\\
Looking for solutions of the form
\beq
\rho(z)  = \rho_0  + f(z),
\label{rhof}
\eeq 
and substituting for the expressions for $M_g$ and $N_g$ 
 into Eq. (\ref{gperhoeqn}), 
we get
\begin{equation}
\frac{\hbar}{2m} \frac{df}{dz} = \pm f [\frac{fg}{m} + c_g^2 \gamma_g^2]
\label{dfdz}
\end{equation}
where 
$c_g = \sqrt{ g \rho_0/m}$
 is the Bogoliubov speed of sound we found earlier 
(see above (\ref{ansatz}), and 
$\gamma_g^2 = 1 -  v^2/c_g ^2$. 
Equation (\ref{dfdz})   can be integrated to give
\begin{equation}
f(z) = - \rho_0\, \gamma_g^{2}\, \rm{sech} ^{2} 
(m/\hbar)  \gamma_g c_g z , 
\label{gf}
\end{equation}
yielding the well-known GP dark soliton solution
\begin{equation}
\rho(z) = \rho_0 [ 1 - \gamma_g^2\, 
\rm {sech}^2 (m/\hbar) \gamma_g c_g z].
\label{grho}
\end{equation}
This is a dark soliton that describes a depression in 
the background density $\rho_0$.
Its profile  flattens out as $v$ tends to the speed of sound $c_g$. 
{\em Thus the GP- dark soliton is of NP-type.}

The phase of the soliton is obtained by substituting 
Eq. (\ref{grho}) into
Eq. (\ref{gradphi}) and   integrating it to give
$\phi(z) = -\tan^{-1}[ (\gamma_g c_g/v) \tanh \gamma_g c_g z]$.
This yields the phase jump across the soliton to be
$ \Delta \phi =  -2 \cos ^{-1}[\frac{v}{c_g}]$.

 It is important to note that while we looked for solutions
for $\rho$ and $\phi$ as in (\ref{ansatz}) that had
  two parameters $v$ and $\omega$,
 Eq. (\ref{gomega}) shows that for the GP soliton,
frequency $\omega = \omega_{g} = -g \rho_0/\hbar$  
is not a variable  parameter, but is determined  
by the  local repulsion energy  $g$ 
and the background condensate  density
$\rho_0$. Thus the GP soliton has only a 
single variable parameter, its 
 speed $v$ which cannot exceed the speed of sound.

Further, an inspection of Eq. (\ref{gpe}) shows that 
$\omega_g$ is 
has its origin in the purely time-dependent phase 
$\phi = \omega_g \tau$ associated with the  background
condensate density $\rho_0$, which is nonzero for the GP soliton.
In other words,  $\hbar \omega_g$
can be regarded as  the energy of the background. 
 
Finally, it is instructive to write $f$    
in terms of the BH model parameters by 
setting $\hbar = m = 1$ and  $g = U$, with U denoting a  
dimensionless  variable $(U/t)$ :

\begin{equation}
f(z) = -\frac{ c_g^2 \gamma_g^2}{U [\cosh  2 \gamma_g c_g z +1]}
\label{gfd}
\end{equation},

\section{ BEC evolution for strongly repulsive bosons and HGPE:
 Order parameter evolution using spin-coherent states}

We are interested in studying  the condensate of a  
strongly repulsive  boson system,
described by the hard core boson limit  $U \rightarrow \infty$.
Firstly, we note that  if we  set  the repulsion $U\rightarrow \infty$,  then since $c_g^{2} \rightarrow U \rightarrow \infty$ and $\gamma_g \rightarrow 1$,  the GP soliton (\ref{gfd}) found from the usual BH  model
 flattens out and delocalizes. As we shall see,  the addition  
 of a nn attraction $V$  as in the extended BH  model (\ref{BH})  helps in  localizing  the soliton in the HCB system.
  
\subsection {HCB system : Mapping to  Spin- $\textstyle \frac{1}{2}$
 Hamiltonian}
The limit $U \rightarrow \infty$ in the Hamiltonian (\ref{BH})
implies that
 two bosons  cannot occupy 
the same site. i.e., boson operators anticommute at same site 
but  commute  at different sites. This leads to 
$ b_{j}^{2} =0 \,;\, n_{j}^{2} = n_{j}\,;\,\,\{b_{j},\,b_{j}^{\dagger}\} = 1 \,\,\,; \,\,[b_{j},\, b_{l}^{\dagger}] =(1 -2 n_j) \delta_{j l}
$. 
Identifying $b_j$ = $S_{j}^{+}$  (spin-raising operator) and
$n_{j} = \frac{1}{2} - S_{j}^{z}$ (operator for $z$-component of spin) 
yields the
spin-$\frac{1}{2}$ algebra:  $[S_{j}^{+},\, S_{l}^{-}] = 
2 S_{j}^{z}\, \delta_{jl}$.
Using the above identification to spin operators, the 
extended Bose-Hubbard Hamiltonian (\ref{BH}) 
for HCB  maps to the following 
quantum XXZ Heisenberg spin-$\textstyle \frac{1}{2}$  ferromagnetic 
(since $t > 0$)  
Hamiltonian in a magnetic field along the $z$-axis: 
\beq
H_S=-\sum_{j, a}[t\,(S_j^+S_{j+a}^-+h.c) +V S_j^z S_{j+a}^z]-
\sum_j((t -V)d)S_j^z.
\label{XXZ}
\eeq

\subsection { Order parameter evolution for the HCB system: HGPE}

The dynamics of the HCB system is given by the Heisenberg  equation
of motion:
\begin{widetext}
\beq
i \hbar \dot{S}_j^+  =[ S_j^{+}, H_{S}]
=  (t - V)d~~
S_j^+ - t S_j^z\sum_{a}S_{j+a}^+ + V S_j^+ \sum_{a} S_{j+a}^z .
\label{SD}
\eeq
\end{widetext}
 Since the condensate order parameter is the expectation
value of the boson operator in a system, it is easy to see that 
for the hard-core boson system 
 it becomes $\eta_j$, the expectation value of the spin-flip operator ,
 i.e.,  $\eta_j=\langle S_j^+\rangle$.

 We use spin-coherent states \cite{radcliffe} as the natural choice for 
computing the above expectation value, due to the  inherent 
coherence in the condensed phase of the HCB system \cite{ourprl}.
This is analogous to the use of boson coherent states 
for defining the order parameter of a  {\em weakly repulsive} system 
of normal bosons,  which leads  to  the GPE, as we saw above. 

The spin coherent state at a lattice site $l$ is defined by
$|\tau_{l}\rangle =(1 +|\tau_{l}|^{2})~~\exp~~ [\tau_{l}\,S_{l}^{-}]~~
|0\rangle $,
where  $S_{l}^{-}= S_{l}^x - i S_{l}^{y}$ is the spin lowering operator,
 $\tau_{l}$ is a complex  quantity, and
 $S_{l}^{z} |S\rangle = S |0\rangle$. 
For $N$ spins, we work with the direct product
 $|\tau\rangle = \Pi _{l}^{N} |\tau_{l}\rangle $.
The states  $|\tau_{l}\rangle$ are normalized, 
nonorthogonal and over complete.
It can be shown that  
the diagonal matrix elements of   {\it single} spin 
operators in the spin coherent 
representation are identical to the corresponding expressions for a 
classical spin \cite{radcliffe}.
 For $S = \frac{1}{2}$,  it can be shown that
the  condensate number density  $\rho^s_j=|\eta_j|^2$
and the
 particle number density $\rho_j =\langle n_j\rangle$ are related 
 by \cite{ourprl} 
\beq 
\rho^s_j=\rho_j (1-\rho_j)  = \rho_{j}\,\, \rho^{h}_{j}, 
\label{rhos}
\eeq
 where $\rho^{h}_{j} = (1 - \rho_{j})$  is the hole density.
 Hence both particles and holes play equally 
important roles in determining the condensate properties.
Further, in contrast to the GPE case, $\rho^{s} \ne \rho$, 
 implying the presence of depletion in the HCB system.

As explained in \cite{ourprl}, taking the  
spin coherent state expectation value of Eq. (\ref{SD}) leads to 
 the evolution equation for the order parameter $\eta_j =<S_j^{+}>$ 
  on the lattice.   
A continuum description  
of the discrete equations is useful 
  when the order parameter is a  smoothly varying
function with  a length scale greater than the lattice spacing $a$.
Using appropriate Taylor expansions for the various  quantities 
appearing  in the lattice equations \cite{ourprl} , we get   
\begin{equation}
i \hbar\, \frac{\partial{\eta}}{d \tau}
= -\frac{t a^2}{2} (1 - 2\rho) \,\,\,\nabla^2 \eta 
-V a^{2} \,\,\,\nabla^2  \rho ~~\,\, \eta 
+ 2 (t - V) d~~ \rho~~ \eta, \label{hgpe}  
\end{equation}
where $\tau$ stands for time and $d$ is the dimensionality
of the lattice. We call this equation  HGPE, "H" representing HCB.
Note that in Eq. (\ref{hgpe}), the condensate wave function is given by
\begin{widetext}
\begin{equation}
\eta({\bf r},\tau) = \sqrt{\rho^s ({\bf r},\tau)} \exp (i \phi({\bf r},\tau)) 
= \sqrt {\rho({\bf r},\tau) (1 - \rho ({\bf r},\tau)) } 
\exp (i \phi ({\bf r},\tau)), \label{etar}
\end{equation}
\end{widetext}
where we have used Eq. (\ref{rhos}).
Substituting Eq. (\ref{etar}) into Eq. (\ref{hgpe}), 
coupled nonlinear evolution equations for the particle 
density $\rho$ and the phase $\phi$ can be written down. From their
solution, the condensate density  $\rho_s = \rho(1-\rho)$  as well as
the condensate wave function $\eta$ can  be found.  

While our discussion so far is for d-dimensions, 
our interest in this paper
is  to investigate solitons in a BEC trapped in a one-dimensional lattice/ 
highly anisotropic, cigar-shaped trap \cite{daux}. 
Therefore in what follows, we will set $d =1$, 
and look for unidirectional traveling wave solutions.

\section {Some general features of HGPE}
Before proceeding to our analysis of soliton
solutions of HGPE, we point out some {\em general} 
features of  the condensate 
parameter evolution of the HCB system as described by 
HGPE (Eq. (\ref{hgpe}). These will be useful
in understanding the various characteristics of 
the soliton solutions
 we will find for this system. 
\subsection {GPE as a certain low-density  approximation 
to HGPE}
From Eq. (\ref{rhos}), we note that in the low density approximation, 
we can set $\rho^s \approx \rho$. Using this  in Eq. (\ref{etar}),
we have, $\eta \rightarrow \psi = \sqrt{\rho} \exp{ i \phi}$. 
In addition, if we  also {\it neglect}
nonlinear terms involving 
$\rho~~\nabla ^2{\eta}$ and $\nabla^2{\rho}~~\eta$
in Eq. (\ref{hgpe}), we get the GPE given in Eq. (\ref{gpe}),
but with the identification  
$t a^{2} = \hbar^2/m$,  and with  $2 (t -V)$  as an
effective local interaction between  the (hard-core) bosons
in the GPE limit.
While  in Eq. (\ref{gpe}) obtained from the 
condensate dynamics of the usual BH model with normal bosons
 is always repulsive 
 ( $g > 0$), 
$2 (t -V)$ arising as an approximation
to the condensate dynamics of the  extended BH model
for HCB can be positive, negative or zero. However, by comparison with
 GPE discussion of $c_{g}$,  the sound speed  will be given by 
 $ \sqrt{2 (t - V) \rho_{0}/m}$,
for the  approximated HGPE under consideration . Hence, only for  
$(t - V) > 0$, 
will there be a speed of sound for this limit.
\subsection {Particle-hole symmetry}
 In the  HGPE (Eq. (\ref{hgpe})), if we set 
$(1 - 2 \rho) =  (\rho_h - \rho)~;~  
\rho  =  [1 +  (\rho - \rho_h)]/2$,
we get
\begin{widetext}
\beq
i \hbar\, \frac{\partial{\eta}}{\partial \tau} 
= -\frac{ta^2}{2} (\rho_h  -  \rho) \,\,\,\eta_{xx}  
-\frac{V}{2} a^{2}  \,\,(\rho - \rho_{h})_{xx}\,\, \eta
+  (t - V) \,(\rho - \rho_h)\,\, \eta + \,\,(t - V)  \eta \\ 
\label{hg-ph}
\eeq
\end{widetext}
We may use the gauge transformation 
  to remove the last term in Eq. (\ref{hg-ph})
\begin{widetext}
\beq
\eta \rightarrow 
\eta \exp - i (t - V)) \tau/\hbar,
\label{gt}
\eeq
\beq
i \hbar\, \frac{\partial{\eta}}{\partial \tau} 
= -\frac{ta^2}{2} (\rho_h  -  \rho) \,\,\,\eta_{xx}  
-{\frac{V}{2}} a^{2}  \,\,(\rho - \rho_{h})_{xx}\,\, \eta
+  (t - V) \,(\rho - \rho_h)\,\, \eta  
\label{hg-eta}
\eeq
\end{widetext}
In the above equation,  we observe that interchanging
the particle density $\rho$ and the hole density $\rho_h$ 
  changes the 
overall sign of the right hand side. Also, $\eta$ remains invariant 
under this interchange. This  shows that if  
  $\eta$  is the 
 wave function for the condensate of particles, $\eta^{*}$  becomes 
the  wave function for the condensate of holes. Thus   
Eq. (\ref{hg-eta}) has a
particle-hole symmetry and  proves to be convenient   
for obtaining a unified formulation of the HCB 
condensate dynamics that we seek.  
Rewriting Eq. (\ref{hg-eta})  in terms of $\rho$ alone,
 we obtain 
\beq
i \hbar\, \frac{\partial{\eta}}{\partial \tau} 
= -\frac{ta^2}{2} (1 - 2 \rho) \,\,\,\eta_{xx}  
-V a^{2}  \,\,\rho_{xx}\,\, \eta
+  (t - V) \,(2 \rho - 1) \,\, \eta . \\ 
\label{hgpe-gt}
\eeq

\subsection { Fractional filling and integer filling 
background densities: Differences in physical characteristics}

 One usually looks for  solutions for the condensate 
that are spatially homogeneous asymptotically,i.e., 
$\rho \rightarrow \rho_{0}$, $\eta \rightarrow \sqrt{\rho_0 (1 -\rho_0)} \exp (i \phi_{\pm \infty})$. On substituting
this  asymptotic solution into Eq. (\ref{hgpe-gt}), we find that for 
fractional filling backgrounds  $0 < \rho_0 < 1$ for which 
the  condensate background density is nonzero asymptotically, the phase must have 
a  purely time dependent term $ \omega_{F} \tau$ as well, 
with intrinsic frequency determined in terms of  system parameters as
$\hbar\omega_F/t  = E_e [1 - 2\rho_0]/\hbar$,
where $E_e$ is a dimensionless  effective energy parameter
\beq
E_e = \frac{t - V}{t},
\label{E2}
\eeq

and the subscript $F$ denotes fractional filling.
In contrast, for the integer filling background with 
$\rho_0 =0$ or $1$ which
implies a vanishing condensate density,  Eq. (\ref{hgpe-gt}) is identically
satisfied asymptotically, and  hence the frequency
$\omega$ does not get determined, and 
 remains a variable parameter.

Secondly,  linear excitations of the HGPE analyzed 
using  small amplitude solutions of Eq. (\ref{hgpe})  
yields the speed of sound  in the HCB condensate as \cite{ourprl}
\beq
 c  \sim  \sqrt{2 E_e  \rho_{0} ( 1 - \rho_0)},
\label{chg}
\eeq
 This implies that while there are  
sound wave modes for the fractional filling 
background when $E_e> 0$,  they are absent  for the integer filling case. 
For  $ E_e \le 0$,
the HCB system does not support sound waves for any filling.

Consistent with the above  observations, we will indeed 
find that  soliton solutions with  fractional filling and integer 
filling background densities  belong to two distinct classes,
with only the former getting associated with a fixed 
frequency $\omega_{F}$ given earlier in this subsection,
and a speed of sound  as in  
 Eq. (\ref{chg}).\\

\section {Soliton solutions for the HGPE }

Setting $\eta = \rho ( 1 - \rho)$ in Eq. (\ref{hgpe-gt}) and equating real and imaginary parts, we obtain the following coupled equations  for
$\rho$ and $\phi$:
\begin{widetext}
\beq
\hbar \rho_{\tau}/t  = - a^{2} [\rho ( 1 - \rho) \phi_{x}]_x
\label{rhotau}
\eeq
\beq
\hbar \phi_{\tau}/t =  E_{e}  (1 - 2 \rho) + \frac{a^2}{4} \big[ \frac{\rho_{xx}}{\rho ( 1- \rho)} - \frac{(1 - 2 \rho) \rho_x^2}{2 \rho^2 (1 - \rho)^2}\big]
- E_{e} a^2 \rho_{xx} -\frac{ a^2}{2} (1 - 2 \rho) \phi_x^2
\label{phitau}
\eeq
\end{widetext}

\subsection {Exact nonlinear plane wave solutions}

Although our interest is in finding    soliton solutions which are localized in space, it is interesting to note that  the above coupled nonlinear PDEs  have {\em exact} plane wave solutions. Taking
$\rho (x, \tau) = \rho_0$, Eq. (\ref{rhotau})  gives $\phi_{xx} = 0$. This  leads to  plane wave solutions
 $\phi (x, \tau) =
-k x  + \Omega \tau$. Using  this in Eq. (\ref{phitau}), we get 
 the exact dispersion relation for the plane waves, quadratic in $k$ :
\beq
\hbar \Omega (k)/t  = ( \hbar \omega_{F}/t)  -  (\frac{1}{2} -  \rho_0) a^2 k^2
\label{plane}
\eeq 
where $\omega_{F}$ is the same as that found in the previous section. 
Thus for  $\rho_{0} > \textstyle\frac{1}{2}$,
the plane wave excitation is like a particle, whereas for
$\rho_0 < \textstyle \frac{1}{2}$  it  is hole-like.  For 
 $\rho_{0} = \textstyle{\frac{1}{2}}$, $\Omega(k)$ vanishes, showing that the 
plane wave becomes static.

\subsection{Solitons for fractional and integer filling background : 
A unified formulation}

While the methodology  that we will use to find solitary wave solutions of HGPE will be in close parallel with that of the GPE 
discussed in the last section, HGPE
will be seen to support both bright and dark solitons, in contrast 
to the GPE which has only dark solitons. This essentially 
 arises due to a particle-hole symmetry
in the HCB model.

Looking for unidirectional traveling waves of the 
typical form (\ref{ansatz}), Eqs. (\ref{rhotau}) and (\ref{phitau})  become

\beq
v \rho_{z} =  \big[\rho (1 -\rho) \phi_{z}\big]_{z}
\label{hrhoz}
\eeq
and
\begin{widetext}
\beq
[E_{\omega} - v \phi_{z}]  \rho_{z}= E_{e} (1 -2\rho) \rho_{z}+ 
\frac{1}{8} \frac{d}{dz}[\frac{\rho_z^2}{\rho(1 -\rho)}] - E_{e} \rho_{zz} \rho_{z}
-\frac{1}{2} (1 - 2\rho) \phi_{z}^2 \rho_{z}
\label{hphiz}
\eeq
\end{widetext}
where $v\hbar/a t$  has now been defined as a dimensionless speed $v$.
$E_{\omega}$ is 
a  dimensionless energy (in units of hopping energy $t$) defined  by
\beq
E_{\omega} =  \frac {\hbar \omega}{t},
\label{Eom}
\eeq
Using boundary conditions $\rho \rightarrow \rho_0$ and 
$\phi_{z} \rightarrow 0 $ as $|z| \rightarrow \infty$, 
Eq. (\ref{hrhoz}) can be easily integrated to give
\beq
\phi_{z} =   \frac{v(\rho - \rho_0)}{\rho ( 1 - \rho)},
\label{hphiz1}
\eeq
Substituting Eq. (\ref{hphiz1}) in Eq. (\ref{hphiz}) and integrating,
we get the following general nonlinear ordinary differential equation  
valid for all values of $E_e$ and 
background densities $\rho_0$:   

\begin{widetext}
\beq
\frac{1}{4} (d\rho/dz)^{2}\,[ 1 - 4 E_e \rho (1 - \rho)] =
  [-\rho_{0}^2 v^{2} +  (2 \rho_{0} - 1) v ^{2}- \lambda_0] \rho
\,+ \,[2 (E_e - E_{\omega})- \lambda_0] \rho^{2} + 2 (2E_e - E_{\omega}) \rho^{3} - 2 E_e \rho^{4}]
\label{quartic}
\eeq
\end{widetext}
where  $\lambda_{0}$ is an integration constant.  
It is interesting to note the natural appearance of 
the  two  energies $E_{\omega}$ and $E_e$, 
associated with the frequency parameter $\omega$ in the phase $\phi$,
and an effective energy parameter $(t - V)$ in the 
BH model for HCB, respectively.
 
From  Eq. (\ref{quartic}) we see 
that  $(d\rho/dz)^{2}$ 
can be approximated to a quartic polynomial for small values of $E_e$. 
Since we are interested in finding
localized solutions for $\rho (z)$, with the asymptotic
boundary condition $ (d\rho/dz) \rightarrow 0$ 
and $\rho \rightarrow \rho_0$
as $z \rightarrow \pm\infty$,
 the quartic polynomial on the right hand side of
Eq. (\ref {quartic}) can be written in the form
 
\begin{equation}
\frac{1}{4} (d \rho/dz)^2 = (\rho-\rho_0)^2 [L \rho^2 + M \rho + N].,
\label{rhoz1}
\end{equation}
where the unknown quantities $L,  M,  N $ are 
to be found  consistently by equating  the 
$\rho^n$ terms ($n = 0$ to $4$) on 
the right hand sides of Eqs. (\ref{quartic}) 
and Eq. (\ref{rhoz1}).  Although this is a straightforward 
analysis, we give some details to show how the 
difference between the fractional and integer filling cases arises.
We get
\begin{widetext}
\beq
L = -2 E_e\label{L} ;\,\, M = 2 [ E_{e} (1 -\rho_{0}) - E_{\omega});\,\, 
N = -2 (E_{e} - E_{\omega}) ~\delta (\rho_{0}) - v^{2},
\label{paraH}
\eeq
\end{widetext}
 where for convenience we use the notation  $\delta (X) = 1 (0)$ for 
$X = 0  (X \ne 0)$. 
 
We also get the following consistency condition:
\beq
(1 -2 \rho_0) (N + v^{2}) + 2 (1 -\rho_0)^2 [E_{e} (1 - 2  \rho_{0}) - E_{\omega}] = 0
\label{N2}
\eeq  
 From Eqs. (\ref{paraH}), it is easy to see that for 
the integer filling cases  $\rho_0 = 0$ and $1$ , 
Eq. (\ref{N2}) is identically satisfied. In contrast,
for the fractional filling case
 $0 < \rho_0 < 1$, since  
from  Eq. (\ref{paraH}), 
$N_{F} = -v^2$,  Eq. (\ref{N2}), leads to the following 
 {\em constraint} on $E_{\omega}$
\beq
E_{\omega} =  E_e ( 1 - 2\rho_0) = E_{\omega_{F}}.
\label{con}
\eeq
This  shows that for fractional filling, the two energies get related, 
so that the frequency  takes on
 the {\it fixed}   value
\beq
 \hbar \omega_{F}/t  =  E_e  (1 -2\rho_0 ),
\label{OmF}
\eeq
which is {\em dependent} on the effective energy of the HCB system. 
This is also in agreement with the expression found in the general features 
of the system discussed in the previous section (see discussion 
above Eq. (\ref{chg})).
In  contrast, for integer filling  $\rho_0 =0,1$, 
the frequency  $\omega$ is 
{\em not} determined, and  is hence an independently   
{\em variable}  parameter.

Eq. (\ref{OmF})  yields 
$\omega_{F} (\rho_0) = -\omega_{F} [(1 - \rho_0)] = - \omega_{F} (\rho_{h})$
showing the particle-hole symmetry explicitly, 
for fractional $\rho_{0}$.  
We remark that $\lambda_0$ can also be found consistently.
Further, the  analysis presented above is valid for small $E_e$ 
positive, negative or zero
(See above Eq. (\ref{rhoz1})). 
 For the BH model parameters appearing in Eq. (\ref{BH}), this implies
that $V$ has to be an attractive interaction (i.e., $V > 0$). 

We look for solutions 
\beq
\rho(z) = \rho_0 + f(z).
\label{Rf}
\eeq

Using this in Eq. (\ref{rhoz1}), and substituting for $L,M$ and $N$ from (\ref{paraH}) we see that for all $\rho_0$, we can write   
\beq 
\frac{1}{4} (df/dz)^{2} = f^{2} [ A f^2 + 
2B f + D],
\label{genfz}
\eeq
where the constraint Eq. (\ref{con})  applies for fractional
filling only.  By combining the results for fractional and integer 
filling densities, it is possible to
 write the following  expressions for $A$, $B$ and $D$ 
which are valid for both types of fillings.
\begin{widetext}
\beq
A = -2E_e ;~~
B = E_{e}(1-2\rho_0)\delta(\rho_0-F) + (2E_{e} -  E_{\omega})\delta(\rho_0)
 - (2E_{e} +  E_{\omega} )  \delta(\rho_0-1),
\label{AB}
\eeq
\end{widetext}
where in $B$ given above, $F$ stands for any fractional value, 
$0 < F < 1$, and  $E_{\omega}$ and $E_{e}$
 are defined in Eq. (\ref{Eom}) and (\ref{E2}) respectively.
 \beq
D = (c^2-v^2) = c^2 \gamma^2,
\label{D}
\eeq
where  $c^{2}$ is given by 
\beq
c^2 = 2 E_{e} \rho_0(1-\rho_0) + 2 (E_{\omega} - E_{e})\delta(\rho_0)
+  2(- E_{\omega}  - E_{e} )\delta(\rho_0-1),
\label{ceqn}
\eeq
and $\gamma^2 =  (1 -   \textstyle \frac{v^2}{c^2})$.
Equation (\ref{genfz}) can be solved to give the following 
single functional form for the soliton solution :

\begin{equation}
f^{\pm}(z) = \frac{c^2 \gamma^2}{\pm \sqrt{B^2+  2 E_{e} c^2 \gamma^2} 
\cosh 2c\gamma z - B}.
\label{fsoln}
\end{equation}

\begin{figure}[htbp]
\includegraphics[width = 1.25\linewidth,height=1.25\linewidth]{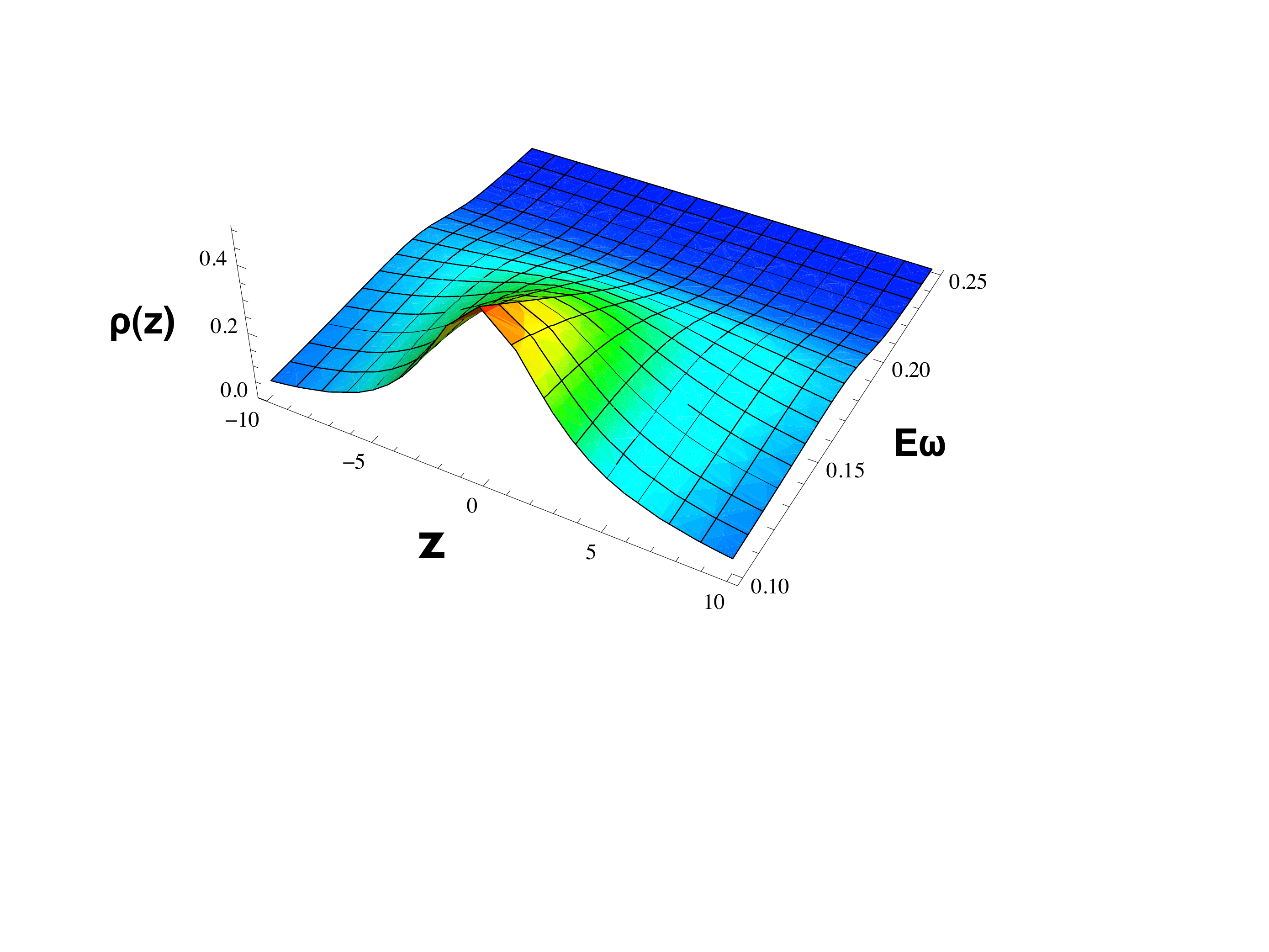}\\
\leavevmode \caption{ Solitons  propagating 
at speed $v/c=.999$ for $\rho_0=0$, $E_e = 0.1$, illustrating 
cross-over from P-type to NP-type with delocalization
for  $E_\omega \ge  0.2$.} 
\label{AA}
\end{figure}

Note that the  soliton solution $ \rho = \rho_{0} + f$  obtained 
from  the single profile (\ref{fsoln})  is valid for {\em both}  
fractional and integer filling backgrounds, 
although these two cases will possess
different {\em physical}  characteristics.  

Due to the following  particle-hole symmetry
\beq
f^\pm (\rho_0, \omega) = -f^\mp\big( (1 -\rho_0), -\omega \big )
\label{ph}
\eeq
it is sufficient to analyze solitons only for the fractional fillings 
$0 < \rho_0 \le \textstyle\frac{1}{2}$ and the integer filling 
$\rho_0 =0$, from which those for $\textstyle \frac{1}{2} < \rho_0 < 1$ and
$\rho_0 = 1$ can be found.

Further, this soliton solution 
is valid for  $E_e =0$, 
as well as  for both signs of $E_e$.  As we will see, 
these  various cases  are quite distinct from each other.

\section{ NP-type and P-type solitons}
For a given $\rho_{0}$,  the two solutions $f^{\pm}$  behave
 differently from each other. 
 In the limit  $\gamma\rightarrow 0$,
 Eq.(\ref{fsoln}) gives
 \beq
 f^{\pm} (v = c)  = \frac{c^2 \gamma^2}{\pm B [1 + c^2 \gamma^2 (E_{e}/B^{2})\,\,+ 2 z^{2}] - B}.
 \label{pmlim}
 \eeq
 This shows that $f^{-}$ vanishes at 
its maximum speed $v = c$. Using 
 $\rho = \rho_0 + f$, this leads to a density soliton
which delocalizes and hence is {\em nonpersistent}  at
its maximum speed. {\em Any soliton with this property will be 
 termed  NP-type soliton.}

Interestingly, at $v =c$, the solution $f^{+}$ tends
 to
 \beq
 f^{+} (v = c)  = \frac{B}{[E_{e} + 2 B^{2} z^{2}]}.
 \label{per}
 \eeq
 Clearly, the corresponding density soliton 
survives and is {\em persistent}  at its maximum speed. {\em Any soliton 
with this property will be termed P-type}.  However, note that its  localized 
 profile vanishes algebraically (rather than exponentially) 
as $|z|\rightarrow  \infty$. 
 The density soliton arising from  (\ref{per})  is P-type and bright for $B>0$ , whereas  it is dark for $B < 0$.  
 
  Further,  the  soliton  becomes  the following  periodic 
soliton train for $v > c$: 
\beq
f^+ ( v > c)  =  2\frac{(v^2-c^2)}
{\sqrt{B^2- 2 E_{e} (v^2-c^2)}\cos 2\sqrt{v^2-c^2} z -B}
\eeq
We will now discuss the characteristics of the soliton for various $E_e$.
\subsection {$E_e = 0$}
For this case, Eq. (\ref{rhoz1}), 
the solution  Eq. (\ref{fsoln}) is exact.  
Secondly, there are no sound waves, 
and {\em no} solutions exist for fractional filling 
backgrounds, $0 <\rho_{0} <1$. Hence we need to 
consider only integer fillings.

(i) $\rho_{0} = 0$: Here, only bright solitons are possible. 
 From Eqs. (\ref{ceqn}) and (\ref{AB}), we get $c^{2} = 2 E_{\omega}$,
hence $\omega > 0$. Hence 
 $B =  - E_{\omega}$ is negative. 
Eq. (\ref{fsoln}) leads to 
\beq
\rho (z) =  \gamma^{2}
\rm {sech}^{2}~(c \, \gamma \, z  )
\label{BRiso}
\eeq
which is a {\it NP-type bright soliton}. 
Here, $c$ is a function of $\omega$.
It is surprising that, 
the bright soliton  (\ref{BRiso})  
for the density  $\rho$ in the strongly 
{\it repulsive} HCB system with a zero background density, 
 has a $\rm {sech}^2$ 
form similar to that of the   
 bright soliton for the weakly 
{\it attractive} GPE. However the detailed characteristics 
of these two solitons are quite distinct. 
 For the latter, the traveling waves 
for the density and phase must travel with different speeds to support a
bright soliton, in contrast with   the former,  
 where they travel with the same speed. 
 In addition, its prefactor is not $\gamma^2$ 
but depends on the above two speeds.

 We remark that bright solitons 
 have  been  predicted and 
observed so far \cite {Bexpt} only  in  weak, locally attractive systems.
Our result (\ref{BRiso}) suggests that these  should be looked for in 
strongly repulsive systems as well.

 (ii) $\rho_0 = 1$: Here, 
 from (\ref{ceqn}), the maximum speed is
$c^{2} = -2 E_{\omega}$ showing that $\omega < 0$.  
Hence $B >0$, yielding NP-type dark soliton of the form 
$\rho = 1 -  \gamma^{2}
\rm {sech}^{2}~(c \, \gamma \, z )$.
While the  form of $\rho$ is exactly the same  as  that   
of  the dark soliton of the  weakly repulsive GPE  with background 
density $\rho_0 = 1$, its  maximum speed is 
{\em not} the  speed of sound.

\begin{figure}[htbp]
\includegraphics[width = 1.15\linewidth,height=1.5\linewidth]{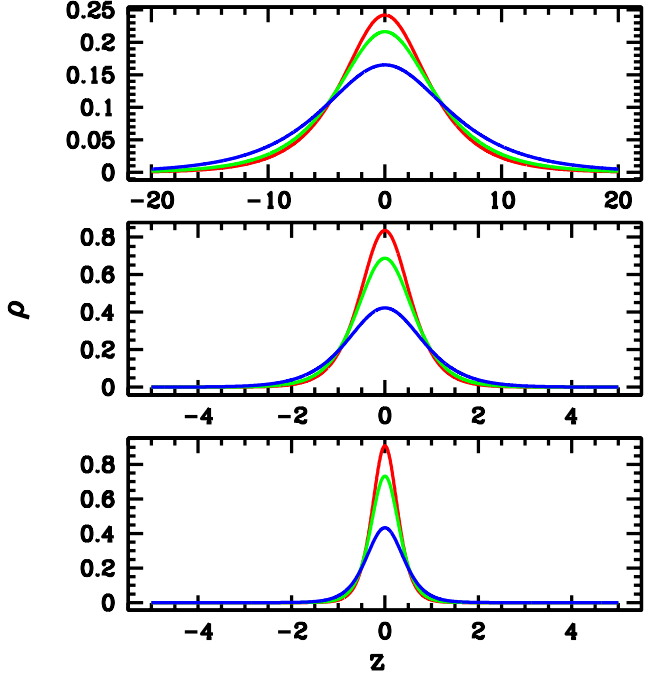}
\leavevmode \caption{ 
For a fixed $E_e= 0.05$, the three  curves
in each plot illustrate bright  soliton profiles for $\rho_0 = 0$,
for $v/c=0.25(red) , 0.5(green),  0.75(blue)$.
The upper plot is for  $c = 0.158$, which is the maximum speed of sound
 possible for the fractional case, when $\rho_0= 0.5$.
The middle and the lower plots are for $c=1$ and $c=2$ respectively. We note that
solitons in middle and lower plots have speeds $6.3$ and $12.6$ times faster
than the speeds in the upper plot. Although $\rho_0=0$ solitons have no
upper limit on their speed, the solitons profile become
extremely narrow as $c$ becomes large.}
\label{ZH}
\end{figure}

\subsection{$E_e > 0$}

(i)  Fractional filling: \\
This case, which corresponds to $0 < \rho_0 < 1$, is the
only case discussed in our previous work \cite{ourprl}.
We summarize the results for this case, in the interest of completeness as well as for
comparison with the other cases to be discussed.
 
  As seen from Eq. (\ref{ceqn}), the maximum speed  of 
the soliton is the speed of sound, 
 $c^{2}  = 2E_e \rho_{0} ( 1 - \rho_{0})$ and 
the parameter $B = E_{e} ( 1 - 2 \rho_{0})$. (See Eq. (\ref{AB}).)
 
  For $\rho_{0} = 1/2$,  since  $B = 0$ , the solutions 
Eq.(\ref{fsoln}) are mirror images of each other. Both of them lead to
density solitons  of NP-type, in the sense that 
they  delocalize as $v \rightarrow c$.
On the other hand, away from half-filling, for 
$0 < \rho_{0} < 1/2$, we have  $B > 0$, showing that
$f^-(\rho_0)$ describes a NP-type dark soliton that 
dies as $ v \rightarrow c$.
In contrast, $f^+(\rho_0)$ leads to a bright soliton on a 
pedestal for the density, that survives 
at $v=c$, i.e., is P-type.  Its exponentially decreasing  profile 
transforms into the following 
algebraic profile:
\beq
f^+( v = c)  =  \frac{1-2\rho_0}{1 + 2 E_{e}(1-2\rho_o)^2  z^2}
\label{alg}
\eeq
 Summarizing, for $0 < \rho_{0} < 1/2$~~ ( $1/2 < \rho_{0} < 1$)
the dark  (bright) soliton is NP-type and dies at  the speed
of sound, whereas
the bright  (dark) soliton is of P- type which survives at  the speed of sound,
taking on an algebraic profile, and becomes a periodic soliton train
at supersonic speeds.\\ 

 (ii) Integer filling $\rho_0 =0, 1$: 

Before discussing this case in detail, in Fig. (\ref{Sall}) we give
 the typical soliton profiles 
for density $\rho$ and condensate density $\rho^s$ found using Eq. (\ref{fsoln}),
for two fractional filling cases as well as the integer 
filling case  $\rho_0 = 0$, for comparison.

Cross-over from P-type to NP-type  soliton: \\

There is an interesting manifestation in the limit $\rho_0 \rightarrow 0$
of the fractional filling case.
In this limit, Eq. (\ref{ceqn}) implies $c \rightarrow 0$, 
so that the soliton becomes {\em static}, with 
a {\em fixed} frequency $\omega_0$ such that $\hbar \omega_0/t  = E_e$.  
But as soon as this limit is reached, the soliton solution 
corresponding to  the integer filling 
$\rho_{0} = 0$  takes over, with $\omega$ becoming a freely variable
parameter. This is discussed below.

 For $\rho_{0} = 0$, firstly, we can have only bright solitons. 
 From Eqs. (\ref{AB}) and (\ref{ceqn}), $B = (2 E_e  - E_\omega)$ 
 and the maximum soliton speed is $c^{2} = 
2 (E_{\omega} - E_e )$. In the $\rho_0 \rightarrow 0$ limit of 
the fractional case, since
 $ \frac{\hbar \omega_{F}}{t} \rightarrow E_{e}$  (see Eq. (\ref{OmF})), this vanishes,
 and the soliton is indeed static. But  $\omega$ is now a variable.
 For real $c$, we must have  $E_{\omega} > E_e$ .
 In addition, as we saw earlier, P-type, persistent bright solitons arise 
 only for $B > 0$,
 i.e., $2E_e \ge E_{\omega}$.
 
 Hence, we conclude that for  $\rho_0 = 0$,  
P-type bright solitons for the density,
 that survive at the maximum soliton speed are supported for the range
 of frequencies $2E_e > E_{\omega} > E_e$.  (Hence  
$\omega$ is positive.) However,  a {\em cross-over} to NP-type  occurs at
  a critical frequency $E_{\omega} = 2 E_e$. In other words,  for  
 all $E_{\omega} \ge 2 E_e$, the solitons are
 NP-type bright solitons that delocalize  at the maximum soliton
 speed. This cross-over phenomenon is illustrated in Fig. (\ref{AA}). 

At the critical frequency  $ E_{\omega} = 2 E_e$,
 $B = 0$.  Using this in Eq. (\ref{fsoln}),
we find the NP-type bright soliton 
\beq
\rho (z)  = \gamma \rm{sech}  2 c \gamma  z,
\label{sech}
\eeq
which delocalizes for  $v = c = \sqrt{2 E_e}$.
Note that both the above expression for the density, as well as that 
for the condensate density 
$\rho^s = \rho ( 1 - \rho) $ 
are  quite different from that of the well known 
bright  GP soliton for GPE with attractive interaction.

For $\rho_0 = 1$, we have only dark solitons. Similar  results 
can be obtained by using particle-hole symmetry.

For a {\em given}  $E_e > 0$, our analysis shows that for the 
fractional filling case, the maximum soliton
speed possible corresponds to half-filling, giving $c^2 = E_e/2$. In contrast, 
for integer filling, it appears as if the maximum speed $c^2 = 2 (E_{\omega} - E_e)$
can keep on increasing as we increase $\omega$. However,
as shown  in Fig. (\ref{ZH}), the width of the soliton keeps decreasing with
$c$. Hence, there will be an effective  speed limit, below which  
the continuum solution we have used will remain valid.

\subsection {$E_{e}  < 0$}

For $E_e \le 0 $, as is clear from Eq. (\ref{ceqn}), 
the system does not support solitons for
fractional background density $\rho_{0} = F$, but can support them 
for $\rho_{0} =0$ and $1$.

(i) $\rho_0 = 0:$ From (\ref{ceqn}), for $c$ to be real, 
$E_{\omega}   > -|E_{e}|$, and
from (\ref{AB}), $B =  [-2 |E_{e}| - E_{\omega}]$.
and the soliton behavior can be found  as follows. 
For $B > 0$, we get 
$E_{\omega} < -2 |E_e|$, which is not consistent with
real $c$. 
Hence {\it no} soliton solutions can arise.
For  $B =0$,
 the solution is purely imaginary, showing that again,`` 
 {\it no} solitons possible.
Finally, for  $B < 0$, we see that 
{\em only} $f^{-}$ solution in (\ref{fsoln}) is possible.
This  leads to  
a NP-type  bright soliton solution 
for  $E_{\omega} > -|E_e|$.

 Hence, for $E_e  < 0$, using particle-hole symmetry relations,
it is easy to infer  that for the background density $\rho_0 = 0$
($\rho_{0} = 1$),
{\it NP-type  bright solitons} ({\it NP-type dark solitons}) 
exist for  
$E_{\omega}  > -|E_e|$~ ($E_{\omega} < |E_e|$), but 
  P-type solitons do not arise at all.\\

This completes the discussion of soliton solutions 
of $\rho (z)= \rho_{0} + f^{\pm}(z)$.  We remark that the 
from  $\rho(z)$, behavior of the solitons for the 
condensate density $\rho^{s}$
can be found. It is also required for the 
discussion of magnetic solitons, as we will
see.

\section {HCB solitons mapped to magnetic solitons in Heisenberg  spin chains}

As pointed out in the beginning, the extended BH model Hamiltonian
for HCB  can be mapped to
the classical  $XXZ$  ferromagnetic Heisenberg spin-$\textstyle\frac{1}{2}$ Hamiltonian, on taking 
spin coherent state average of the Hamiltonian (\ref{XXZ}). The topic
of magnetic solitons in Heisenberg chains that has been studied 
for over two decades \cite{mikes,kose} continues to attract attention
in recent times \cite{NIST} as well. 
While the order parameter for BEC is the condensate 
density $\rho^{s} (z) = \rho(z) (1 - \rho(z))$, the relevant
order parameter for the spin Hamiltonian is $S^{z}(z)$,
which can found from the HCB boson density $\rho(z)$ 
by using the identity $ S^{z} = [(1/2) - \rho(z)]$. 
Thus all the soliton solutions for $\rho(z)$ which we found 
 will also yield the  
corresponding magnetic soliton solutions  for $S^{z}$.

In the last section, we found that in all cases, 
to obtain soliton solutions, 
it is {\em necessary} to have a purely time-dependent 
term $\omega \tau$ in (\ref{ansatz}). For a spin system,
$\omega$ is just the precession frequency due to 
a corresponding ``magnetic field"  
along the $z$-axis.
Thus the  {\em gauge-transformed} 
evolution equation (\ref{hg-eta}) which led to 
 solitons (\ref{fsoln}), also describes those for 
the continuum dynamics of the following 
dimensionless anisotropic Heisenberg spin Hamiltonian:   
\beq
H_{eff}/t=-\sum_{j, a}{\bf S}_j\cdot{\bf S}_{j+a} + 
 E_{e}\sum_{j,a}  S_j^z S_{j+a}^z - E_{\omega} \sum_j S_j^z,
\label {anis-g}
\eeq
where  $E_{e}$  is the strength  of the anisotropy 
energy (see Eq. (\ref{E2})),
and $E_{\omega}$  (see Eq. (\ref{Eom}))  is the appropriate  "magnetic
field" along the $z$-axis, which  
would provide a physical origin for the  above 
necessary  spin precession.  

(i) $E_{e} > 0$: Easy-plane anisotropic chain:

 (a) Fractional background density $0 < \rho_0 < 1$:
Recall that for this case,
which now  corresponds to $(-1/2) < S^{z}_0 < (1/2)$,
this frequency $\omega_{F}$ is {\em fixed} and is given by 
$\hbar \omega_{F}/t  = E_{e} (1 - 2 \rho_{0}) = 2 E_{e}  S^{z}_0$.
 Eq. (\ref {anis-g})  yields 
\begin{widetext}
\beq
H_{eff}/t=-\sum_{j, a}{\bf S}_j\cdot{\bf S}_{j+a} + 
 E_{e}\sum_{j,a}  S_j^z S_{j+a}^z -
\sum_j 2 E_{e} \, S^{z}_0  S_j^z.
\label {Seff}
\eeq
\end{widetext}

 It is to be noted that unlike in usual spin chains, 
 here the "external"  magnetic field  is not an independent 
variable, but depends on the anisotropy $E_{e} > 0$ and $S^{z}_0$.

Two competing terms appear: the easy plane anisotropy tends to 
make spins lie on XY plane, but the "magnetic field" 
 tends to align spins along the $z$ axis.
Thus this magnetic field 
encodes the   particle-hole imbalance 
$(1 - 2 \rho_{0})$) in the background.\\
 For half-filling, since $\rho_{0} = 1/2$, 
"magnetic field"  vanishes.
Hence there is no  preferred 
direction about the $z$ axis, and symmetric  orientations  above and below the 
easy plane are preferred by the excitations, yielding
 solitons  that are {\em mirror images}  of each other (see Fig.(\ref{Sall}) (a)).
 
 Away from half-filling, since $\rho_{0} \ne 1/2$, 
the magnetic field is nonzero, and this symmetry is lost
(see Fig. (\ref{Sall}) (b).
For fractional filling,  $-1/2 < S^{z}_0 < 1/2$,
we  will get both NP-type and P-type magnetic solitons.
These solitons we obtain for spin $S = \textstyle\frac{1}{2}$ are similar to  
the A and B-type rotary wave solutions \cite{kose}
found  for magnetic solitons in spin $S$, easy-plane chains with a 
specific type of external field which depends on the anisotropy and the
boundary condition on $S^z$. 
It is interesting that this type of spin Hamiltonian appears
in a natural fashion in the extended BH model for HCB.

(b) Integer background density $\rho_{0} = 0$ or $1$, corresponding
to $S^{z}_0 = 1/2$ or $-1/2$. Here, the precession frequency $\omega$,
 which is {\em necessary} to create a soliton, is {\em not}  fixed, 
and the spin Hamiltonian is given by  (\ref{anis-g}), with 
the magnetic field term proportional to $\omega$. Translating 
our results for the limit $\rho_0 \rightarrow 0 $ we discussed for BEC,
 we find that it is possible to 
obtain P- type magnetic solitons for $S^{z}$ that persist even at their
maximum speed (which depends on $\omega$) for a range of magnetic
fields $E_{\omega}$, beyond which they {\em cross-over}  
to  NP-type magnetic solitons.

(ii) $E_e = 0$ and $E_e < 0$: Isotropic and Easy-axis anisotropic chain:\\
As is obvious, these can have only boundary conditions $S^{z}_0 = 1/2$
or $- 1/2$, which correspond to integer filling backgrounds $\rho_0 =0$ or $1$. Hence $\omega$ is a variable parameter. However, in contrast 
to the easy plane case, these systems support only NP-type solitons
for all $\omega$.

In the existing  literature \cite{mikes,kose},
magnetic soliton solutions  for $S^{z}$ in 
the classical isotropic chain ($E_e = 0$) as well as the 
easy plane ($E_e > 0$) and the easy axis ($E_e < 0$) 
anisotropic chains have been treated 
individually. As should be obvious, the advantage 
of our unified formulation is that we can now  find 
all of them for various
boundary conditions from a  single functional
form (\ref{fsoln}).

\begin{figure}[htbp]
\includegraphics[width = 1\linewidth,height=1.1\linewidth]{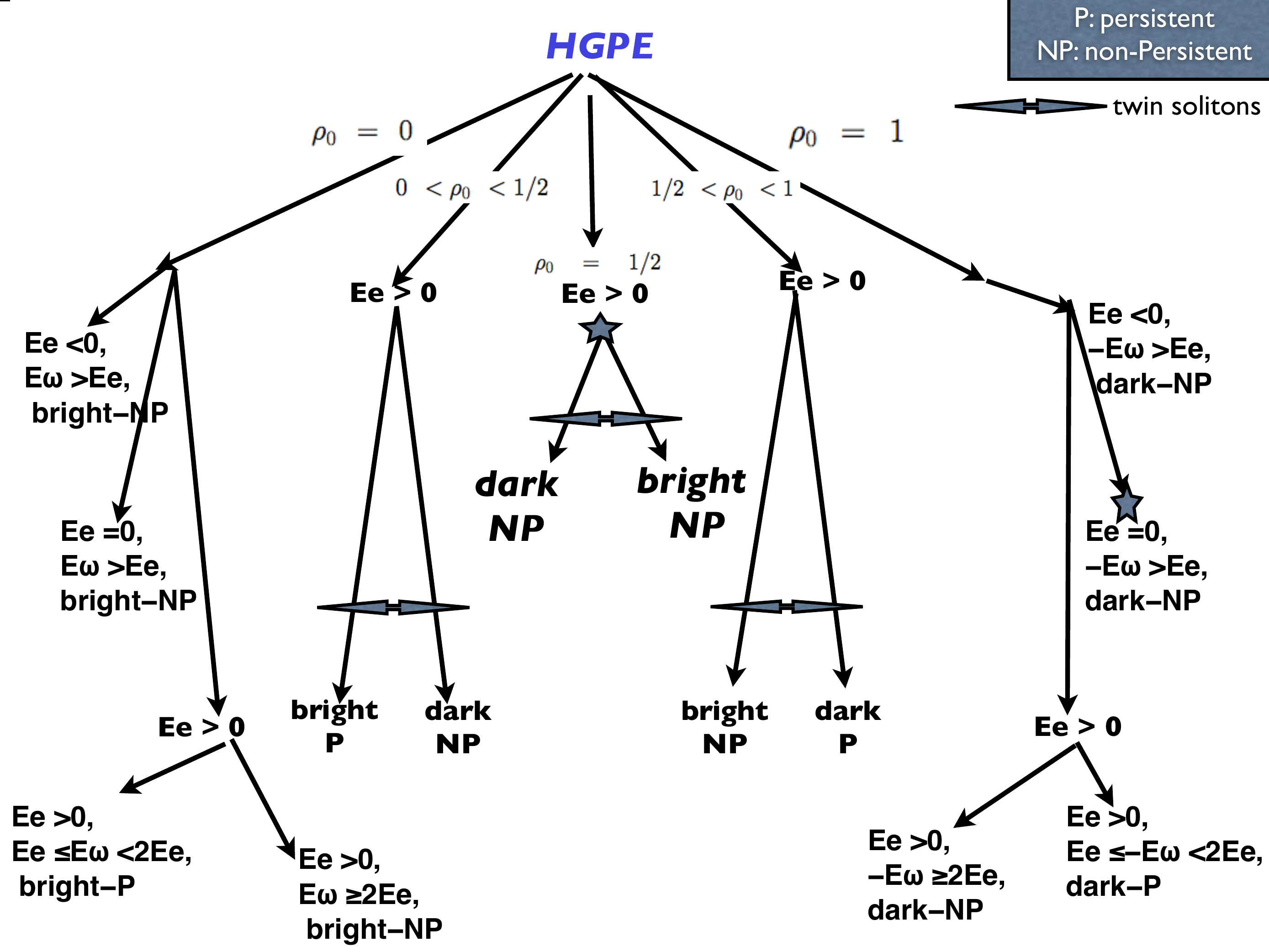}
\leavevmode \caption {Soliton Tree: Classifying the behavior of 
different types of density solitons 
of HCB for both fractional and integer background density $\rho_0$,  
with various  $E_e$ parameters. The star
indicates that the corresponding  soliton  has a form similar to the GP soliton.}
\label{Stree}
\end{figure}

\section{Summary and discussion}

We have provided a unified formulation for finding solitary waves 
with various background densities, in the BEC  of  
strongly repulsive  bosons,  
described by a hard-core boson (HCB) system. Using an extended
Bose-Hubbard (BH) model for HCB, which also includes nearest 
neighbor attractive interactions  on the lattice, we show that
 in the continuum version, the 
 condensate order parameter of this system satisfies Eq. (\ref{hgpe}),
named by us as HGPE. 
 Our comprehensive analysis also includes the GPE for weakly repulsive
 BEC, arising  from the BH model of normal bosons. Interestingly, 
the GPE also emerges 
on neglecting certain nonlinear terms in the HGPE for low densities.

We find that while the {\em infinite} on-site repulsion condition 
$U/t \rightarrow \infty$, (i.e., $t << U$)  for HCB in the BH model 
completely delocalizes 
the dark soliton in the GPE,
 (see Eq. (\ref{gfd})),  the addition
of a  {\em finite} nn attractive potential $V \sim t$  which is 
much smaller than $U$, is sufficient
  to  localize the dark soliton or even support bright solitons. As our results
derived from (\ref{fsoln}) for the HGPE show, the behavior of the soliton depends
on condensate boundary conditions as well as the sign 
and magnitude of $E_e$.

Commonality in the methodology to  find  solitons for
both GPE and HGPE with various  background densities,
 provides a simplified
approach to understand these nonlinear modes in both cases. It also
brings out certain universal aspects of the solutions,
as well as certain distinguishing features.

We find it convenient to work with a {\em gauge-transformed}  HGPE which
has an inherent  particle-hole symmetry. 
Solitary waves with amplitude and width expressed in terms of 
its maximum speed $c$ and the corresponding $\gamma$ 
 (see Eqs. (\ref{gfd}) and (\ref{fsoln}))  depict a kind of universality 
that emerges from our theoretical analysis which treats HGPE for various
background densities  as well as the GPE limit in equal footing. 
In addition to highlighting the universal aspects, we show that
the solitons existing with and without background 
condensate density encode a fundamental distinction.
Although in general, two competing energy scales $E_e$  and $E_{\omega}$ 
appear in the HCB system, our systematic analysis shows that
 for the former  these get related, 
giving a fixed frequency, whereas for the latter  
they remain independent parameters.

When the background condensate density is nonzero 
(i.e., fractional filling $\rho_0$) 
 solitons exist only for  $E_{e}   > 0$, with the speed of sound
as the maximum speed of the soliton. 
The soliton is characterized by its speed $v$ alone, while
its  associated frequency $\omega_{F}$ is a constant, fixed by
the background $\rho_0$ and the system parameters.  
For this case, 
the  two species of condensate density solitons coexist: 
A dark soliton which is NP-type, 
and a novel P-type
which persists even at sound speed, when it  becomes fully  bright.

 When the background condensate density vanishes 
(i.e., integer filling $\rho_0$)
 the condensate density solitons exist  for $E_e > 0$ as well as $E_e \le 0$.
 There is no intrinsic speed of sound, nor a fixed frequency for these.
The soliton is a function of $v$  as well as 
its associated  variable frequency $\omega$. Further,
 the maximum soliton speed depends
  upon $\omega$. 
For $E_e > 0$, the two species do not coexist. With  an additional independent energy
scale $E_{\omega}$ that emerges for the vanishing background density,
the P-type bright solitons that survive even at the  maximum speed
$v = c$ exist only provided
 $ E_e < E_{\omega} < 2 E_e$. This  shows that the energy
$E_{\omega}$ (or "ground state" energy) associated with such a
background  should be at least $E_e$,
to create even a static  soliton excitation above the background.
Energies higher than this make the soliton move,  with the  amplitude
at its {\em  maximum} ($\omega$-dependent) speed $v =c$ remaining finite,
showing its persistent nature. However, the amplitude 
for $v = c$ steadily decreases,
till it  vanishes at a {\em critical frequency} 
$E_{\omega_{c}} = 2 E_{e}$, signaling a {\em cross-over} to the
NP-type soliton for  $E_{\omega} > 2 E_{e}$. 

A novel aspect of solitons in the HCB system described by HGPE is the 
possibility of creating very high speed localized modes in a system
whose background has vanishing condensate density, by increasing the
soliton frequency $\omega$.
The zero-background solitons for this strongly repulsive system 
 can indeed be made
to propagate with speeds that are much higher than the 
possible speeds in cases where solitons move in a 
background with a nonvanishing condensate density.

For $E_e \le 0$,  
 there is only one species of solitons for the condensate density.
It is  the NP -type bright soliton, for  zero background 
condensate density.
 
Finally, by using the relationship $\rho(z) = \textstyle\frac{1}{2} - S^z$,
between  HCB density and spin,
 we are also able 
to discuss  the corresponding class of magnetic solitons in 
 isotropic and anisotropic ferromagnetic spin chains in the presence
of a magnetic field along the $z$-direction, in a unified fashion.

The ``Soliton Tree" diagram given in Fig. (\ref{Stree}) summarizes
various possible solitary wave solutions for the density in the HCB system, 
providing a comprehensive picture of nonlinear modes in this strongly
interacting bosonic system. 
  
Soliton propagation has been studied experimentally in  
BEC using various techniques \cite{frantz} such as the
phase-imprinting method \cite {expt, Bexpt, becker}, which manipulates
the initial BEC phase without affecting its density, 
the density-engineering method \cite {dutton, burger} which creates
an appropriate initial form for the  density 
without affecting the BEC phase,
and the quantum-state engineering method \cite {becker, burger} which manipulates 
both the density and the phase.  
We hope that our theoretical results will motivate experimental
research on BEC solitons in the HCB system  we have studied.

ACKNOWLEDGMENTS:\\
RB thanks the Department of Science and Technology, 
India, for financial support.\\


\begin{thebibliography}{99}
\bibitem {def} Although strictly speaking, a soliton 
is a  solitary wave that retains its characteristics 
intact even on collision, we shall use "soliton" to 
denote a "solitary wave", in this paper.
\end{thebibliography}
\end{document}